\documentclass[12pt,a4paper]{article}
\usepackage{natbib}
\usepackage{tikz}

\usepackage{xcolor}

\colorlet{drkblue}{blue!61.8!black}

\usepackage[pdftex]{hyperref}

\hypersetup{colorlinks}
\hypersetup{ colorlinks,
	linkcolor=drkblue,
	filecolor=green,
	urlcolor=red,
	citecolor=drkblue
}

\usepackage[hyperpageref]{backref}

\usepackage{pgf}
\usepackage{tikz}
\usetikzlibrary{arrows,automata}
\usepackage[latin1]{inputenc}
\usepackage{verbatim}

\usepackage[left=2.5cm,top=3.0cm,bottom=3.5cm,right=3cm]{geometry}

\usepackage{color}
\usepackage{sgame}
\usepackage{amsmath}
\usepackage{amssymb}

\usepackage{theorem}

\theorembodyfont{\rmfamily}

\theorembodyfont{\rmfamily}

\newdimen\dummy
\dummy=\oddsidemargin
\theorembodyfont{\rmfamily}

\usepackage{setspace}

\title{Mechanism Design Approach to School Choice:\\ One versus Many\thanks{I gratefully acknowledge financial support from the Swiss National Science Foundation (SNSF).}}
\date{September 13, 2018}
\author{Battal Do\u{g}an\thanks{\texttt{battal.dogan@bristol.ac.uk}. Department of Economics, University of Bristol.}}

\begin{document}

\maketitle

\begin{abstract}

A vast majority of the school choice literature focuses on designing mechanisms to simultaneously assign students to \textit{many} schools, and employs a ``make it up as you go along'' approach when it comes to each school's admissions policy. An alternative approach is to focus on the admissions policy for \textit{one} school. This is especially relevant for effectively communicating policy objectives such as achieving a diverse student body or implementing affirmative action. I argue that the latter approach is relatively under-examined and deserves more attention in the future. \\

\end{abstract}

\onehalfspacing

Mechanism Design has been especially successful in allocating students into schools. In a seminal study, \cite{as2003} formulate desirable properties for assignment mechanisms, and in light of this desiderata show that some school districts use \textit{deficient} mechanisms and propose \textit{better} alternatives. Since then, economists have been studying assignment mechanisms for many school districts around the world, formulating desirable properties motivated by various policy objectives, and proposing alternative mechanisms.    

This design process has two essential parts, one of which has been well-studied and understood, while the other has been relatively underestimated. The well-studied and understood part is the formulation of desirable properties for assignment mechanisms and figuring out which properties can be satisfied simultaneously and by which mechanisms. Although design objectives may differ from district to district, the reforms have been centered around two objectives: (i) Optimizing student welfare while achieving fairness and (ii) making it safe for the students to report their preferences truthfully. The Student Optimal Stable Mechanism (SOSM), proposed by \cite{as2003}, achieves the two goals and is now in use in many school districts around the world. Economists have been very successful in not only proposing alternative mechanisms, but also in convincing the public and the policy makers about the \textit{desirability} of the formulated properties. It is important to note here that the formulation and successful communication of the desirable properties is very crucial to induce reforms in practice, since these desirable properties of the mechanisms pave the way for the economists and the policy makers to convince the public for the necessity of a change. A very striking example is the case of England. In 2007, the \textit{nationwide} School Admissions Code \textit{prohibited} authorities from using the older mechanism which did not make it safe for the students to report their preferences truthfully. The justification for the reform given by Department for Education and Skills was that the earlier mechanism ``made the system unnecessarily complex to parents'', pointing to a specific desirable property violated by the old mechanism.    

The other, relatively underestimated, part of the design process is endowing each school with a choice rule (or an admission rule). Although student preferences are elicited from the students, endowing each school with a choice rule is an essential part of the design process. In the earlier school choice literature, the main focus has been on assignment problems where each school is already endowed with a priority ordering over students, and the choice rule of a school is to simply admit highest-priority students up to the capacity. Although there are indeed school districts where there are no policy objectives that would preclude us from endowing each school with such a simple choice rule and directly proceeding with the other part of the design process, there are many school districts with additional concerns, which calls for a non-trivial design of an appropriate choice rule for each school. An important example is achieving a diverse student body or affirmative action. Many school districts are concerned with maintaining a diverse student body at each school while assigning students to schools when each student belongs to one of multiple types (based on factors such as gender, socioeconomic status, or ethnicity). In this case, what is a \textit{good} choice rule that reconciles diversity objectives with other objectives such as admitting students with higher test scores? The answer to this question is not evident.

To this end, most of the literature so far has employed a ``make it up as you go along'' approach by keeping the focus on the design of a mechanism to assign students to \textit{many} schools and, as a \textit{detail} of the assignment mechanism, endowed schools with choice rules that guarantee assignments with certain properties. An alternative approach, which has first been employed by \cite{ey2013}, is to separately focus on the choice problem of \textit{one} school, formulate desirable properties of choice rules for \textit{one} school, figure out which properties can be satisfied simultaneously and by which choice rules, and then study implications of using these choice rules in assigning students to \textit{many} schools.          

The second approach has an important advantage over the first approach. In the first approach, the objectives are solely stated as properties of the assignment mechanism, and it may be difficult to understand according to which principals each school admits students. The second approach provides a clear foundation for each school's choice behaviour in the assignment process. This is a very important advantage because for some objectives, such as achieving diversity, it is either \textit{easier} or more \textit{natural} to communicate policies towards achieving such objectives through properties of choice rules for \textit{one} school, rather than properties of mechanisms assigning students to \textit{many} schools. As an example, consider the Boston school district where diversity is an objective, based on neighbourhood boundaries. The policy of the Boston school district is the following: at  each  school  in  Boston,  half  of  the  seats  at  each  school  are  made  open  to  all
applicants, while the other half are reserved seats that prioritize applicants from the local neighborhood, and reserved seats are filled ahead of open seats. Note that the Boston school district describes its diversity policy to the public by referring to the choice rule of one school, rather than referring to the assignment mechanism which is the SOSM where each school is endowed with a choice rule as described above. To further justify why the current policy is a good policy to achieve diversity, a natural way to proceed is to discover desirable properties underlying the choice rules of the schools in Boston, which we address in \cite{lexi}. 

It is also important to understand how properties of choice rules for one school relate to the properties of an assignment mechanism that uses these choice rules. In fact, there are properties of choice rules for one school such that the natural counterparts of the properties may not be satisfied by an assignment mechanism that uses choice rules satisfying the properties. For example, some school districts aim to favor the underrepresented minority students, and in order to achieve that objective, there is a majority-quota at each school and it is required that no school be assigned more majority students than its majority-quota. Suppose that there is a priority ordering over students based on their test scores. Define a majority-quota choice rule for a school as the choice rule that admits students according to priority until the majority-quota is reached and from then on only admits minority students according to priority until the capacity is reached. Such a choice rule satisfies the following property: if the majority-quota is lowered, a chosen minority student still remains to be chosen. Now, when assigning students to many schools, a natural counterpart of the property that we have defined above is the following: if the majority-quota at each school is lowered, no minority student is worse off. However, it is known that this property is not satisfied by the student optimal stable mechanism which endows each school with a majority-quota choice rule; even worse, when majority-quotas at all schools are lowered, it is possible that all the minority students are worse off \citep{k2012}. In other words, a choice rule which unambiguously favors minority students when there is one school may have perverse welfare consequences for minority students when endowed by schools in a school district that uses the student optimal stable mechanism.  

I believe that the field should grow more in the direction of formulating desirable properties for one school, figuring out which properties can be satisfied simultaneously and by which choice rules, and understanding whether these properties translate into desirable properties of an assignment mechanism that endows schools with these choice rules. I believe that this will facilitate the communication of design objectives and pave the way for further school choice reforms.

\bibliographystyle{chicago}
\bibliography{dogan_FOED}

\begin{thebibliography}{}

\bibitem[\protect\citeauthoryear{Abdulkadiro\u{g}lu and
  S\"{o}nmez}{Abdulkadiro\u{g}lu and S\"{o}nmez}{2003}]{as2003}
Abdulkadiro\u{g}lu, A. and T.~S\"{o}nmez (2003).
\newblock School choice: A mechanism design approach.
\newblock {\em American Economic Review\/}~{\em 93}, 729--747.

\bibitem[\protect\citeauthoryear{Do\u{g}an, Do\u{g}an, and Yildiz}{Do\u{g}an
  et~al.}{2017}]{lexi}
Do\u{g}an, B., S.~Do\u{g}an, and K.~Yildiz (2017).
\newblock Lexicographic choice under variable capacity constraints.
\newblock {\em Working Paper Available at SSRN:
  https://ssrn.com/abstract=2886494\/}.

\bibitem[\protect\citeauthoryear{Echenique and Yenmez}{Echenique and
  Yenmez}{2015}]{ey2013}
Echenique, F. and M.~B. Yenmez (2015).
\newblock How to control controlled school choice.
\newblock {\em American Economic Review\/}~{\em 105}, 2679--2694.

\bibitem[\protect\citeauthoryear{Kojima}{Kojima}{2012}]{k2012}
Kojima, F. (2012).
\newblock School choice: Impossibilities for affirmative action.
\newblock {\em Games and Economic Behavior\/}~{\em 75}, 685--693.

\end{thebibliography}

\end{document}